# Elaboration Of Global Quality Standards For Natural And Low Energy Cooling In French Tropical Island Buildings.

F. GARDE, H. BOYER, J.C. GATINA

Université de la Réunion, Faculté des Sciences, Laboratoire de Génie industriel, BP7151, 15 avenue René Cassin, 97 715 Saint-Denis Messag Cedex 9, France - email : garde@univ-reunion.fr.

**Abstract :**

Electric load profiles of tropical islands in developed countries are characterised by morning, midday and evening peaks arising from all year round high power demand in the commercial and residential sectors, due mostly to air conditioning appliances and bad thermal conception of the building.

The work presented in this paper has led to the conception of a global quality standards obtained through optimized bioclimatic urban planning and architectural design, the use of passive cooling architectural components, natural ventilation and energy efficient systems such as solar water heaters.

We evaluated, with the aid of an airflow and thermal building simulation software (CODYRUN), the impact of each technical solution on thermal comfort within the building. These technical solutions have been implemented in 280 new pilot dwelling projects through the year 1996.

# 1. DEMAND SIDE MANAGEMENT IN THE THERMAL CONCEPTION OF BUILDINGS

There are four French overseas Departments (DOM) : two islands are located in the Caribbean (Martinique and Guadeloupe), one situated 400km to the east of Madagasgar in the Indian Ocean (Reunion Island) and the fourth Department is in the North of Brazil (French Guiana). Each experiences a hot climate, tropical and humid in the islands of Guadeloupe, Martinique and Reunion and equatorial in French Guiana.

*Fig 1 :Growth in number of dwellings - Reunion island*

Each year 20,000 dwellings are built in the French overseas Departments. Three quarters of this development is welfare housing. Initially this new housing is constructed without the comfort of air conditioning or hot water which has led to the haphazard installation of instant electrical hot water boilers and badly situated, thought out and maintenanced individual air conditioning systems. The lack of thermal regulations combined with the economic constraints of a tight budget for construction have led to the development of buildings totally unadapted to the tropical climate. The large population increase in the DOM, the rise in living standards, and the decreasing costs of air conditioning appliances constitute a real energetic, economic and environmental problem.

*Fig. 2 : Daily electric profile - Reunion Island*

*Fig. 3 : Annual evolution of power capacity and electric consumption - Reunion Island*

The above factors result in a high demand for electricity at peak times (see fig. 2-3) as well as the daily and nightly bad management of consumption which has a direct effect on the size of the electrical production plants and therefore future investments.

More of that, in an insular position (such as the case of the DOM) the electrical production is principally that generated by the low efficient burning of fossil fuels which results in high CO and $SO_2$ emissions. Reducing electrical production also means reducing of pollutant emissions.

Considering the economical aspects, the high cost of production also generates a constant high deficite for EDF in the DOM (more than 2 thousand million French Francs in 1995) as the average production cost per kWh is greater than the selling price (the selling price being the same as in mainland France).

All these factors show us that demand side management in thermal conception of buildings is therefore of great economical, social and environmental importance.

A long term overall programme to improve comfort and energy performance in residential and commercial buildings is actually under-way in the overseas departements. In the new housing sector, a quality standard seal has been launched concerning the building structure, the hot water production systems and the air conditioning appliances.

## 2. THE ECODOM STANDARD

This DSM pilot initiative was launched in early 1995 in the French islands of Guadeloupe and Reunion through a partnership between the French electricity board (EDF), institutions involved in energy saving and environmental conservation (ADEME) and construction quality improvment, the ministries of Housing, Industry and the French Overseas Departement, the University of Reunion island and several other public and private partners, such as low cost housing institutions, architects, energy consultants, etc...The objectives are initially to implement the standard to 280 pilot new dwelling projects throughout the year 1996, then, to expand this pilot phase in the residential sector on a much broader scale (2000 new dwellings per year), and complete similar global energy efficiency projects in existing housing and large and medium size commercial buildings.

*2.1 The objectives*

The ECODOM standard aims to simplify the creation of naturally ventilated comfortable dwellings whilst avoiding the usual necessity of a powered air cooling system consuming electricity. ECODOM has both social and economic objectives as it aims to improve thermal quality standards and decrease energy consumption in the housing concerned. The aim of ECODOM is to provide simple technical solutions, at an affordable price, rather than research the ideal bioclimatic building, which is economically and financially unfeasable. Also the simplicity of the technical solutions provided could enable the setting up of something similar in different countries experiencing the same climate. There also exists the work of Malama [1],

Olusmbo [2] and Ratnaweera [3] who have worked on the conception of buildings adapted to a defined climate, in Zambia, Nigeria and Sri Lanka respectively. The work of Matthews for low cost dwellings in South Africa [4] is very closed to the aims of the ECODOM standard too.

*2.2 The prescriptions*

The level of comfort is reached by an architectural conception of buildings adapted to the local climate: the dwelling is protected from the negative climatic parameters (the sun) and favours the positive climatic factors (the wind).

The achievement of a good level of thermal comfort necessitates the application of a certain number of compulsary rules. These prescriptions concern the dwellings immediate surroundings and its constituant components. They cover five points :

1) Position on site (vegetation around the building)

2) Solar protection (roof, walls, windows)

3) Natural ventilation (exploitation of trade winds, and optimized ratio of inside/outside air-permeability of the dwelling) or mechanical ventilation (air fans).

4) Domestic hot water production (servo-controlled night electric drum, sized according to requirements, solar or gas water heaters).

5) Option, air conditioned bedrooms (closed room and efficient, regulated appliances).

### 3. METHODOLOGY

To reach these quality standards, an important number of simulations were computed on each component of the building in order to quantify the thermal and energetic impact of each technical solution on the thermal comfort within the building. Various authors have already worked on specified problems concerning the outside structure of the building : Bansal [5] on the effect of external colour, Malama [1] on passive cooling strategies for roof and walls, Rousseau [6] on the effect of natural ventilation, De Walls [7] for global considerations on the building adapted for a defined climate.

Our approach consisted of the study of typical dwellings, with the use of a building thermal-airflow simulation software. The simulations were carried out on the constituant componants (roof, walls, windows) and on natural ventilation, in a way to estimate the influence of each of the above prescriptions, in terms of thermal comfort and energetic performances. These simulations, their analysis and the synthesis of the results have been presented in a research document [8], available from the authors. In the following paper, we will illustrate the methodology adopted, present the results obtained concerning natural ventilation and present a synthesis of the results for the overall standard prescriptions.

*3.1 The computer program :*

The software *CODYRUN* which we used for our simulations, has already been covered in various publications[9 - 12]. The multiple model aspects are detailed in [10]. Paper [11] deals with the thermal model and paper [12] presents the data structuration and the description of the front end. We will however present the general lay out of this software. One of the advantages of *CODYRUN* is that the software is an efficient building thermal simulation tool, including research and conception aspects, and taking into consideration different types of climate. More precisely, it is a multizone software integrating both natural ventilation and moisture transfers.The choice of the software is thus justified, by the fact that in a humid tropical climate, the building is an open system where the airflow transfer exchanges are very important and the climatic sollicitations are variable. CODYRUN matches perfectly with these objectives as it was designed for that purpose.The computer program has been validated through experiments carried out, on real pilot sites, in a humid tropical climate [13].

When considering the thermal behaviour of a building, its thermal state is determined by the continuous field of temperatures, concerning all points included within the physical limits of the building. The constitution of a reduced model with a finite number of temperatures, is possible by assuming some simplifications (monodimensionnal heat conduction, well mixed volumes, linearized superficial exchanges, ect...).

In relation with the calculation program, the software executes at each step the recognition of airflow patterns, temperature field and specific humidity of each zone.

Based on the nodal analysis, the thermal model relies on INSA's previous simulation code, *CODYBA* [14], and is the main part of the software. With the usual physical assumptions, we use the technique of nodal discretisation of the space variable by finite difference. In addition, the mass of air inside one zone is represented by a single thermal capacity. Thus, for a given zone, the principle of energy conservation applied to each concerned wall node, associated with the sensible balance of the air volume, constitute a set of equations, that can be condensed in a matricial form .

$$[C] \frac{dT}{dt} = [A] T + B \quad (1)$$

At each step, the resolution of equation (1) uses an implicit finite difference procedure and the coupling iterations between the different zones make it possible to calculate the evolution of temperatures, as well as those of sensible powers needed in case of air conditioning. The zone coupling approach of CODYRUN is quite similar to the one used in ESP [15].The most simplified airflow model considers, as known, the airflow rates between all zones. The more detailed model is an airflow pressure model which takes into account the driving effects of the wind and the thermal buoyancy. The problem of large openings in this pressure model is solved with the use of the Walton model. In comparison with other programs, we can say that the airflow pressure model, integrating large openings, is quite similar to TARP [16]. The building is also represented as a network of pressure nodes, connected by non-linear equations giving the flows as a function of the pressure difference. This detailed airflow calculation goes through the iterative solution of the system of non linear equations made up with the air mass conservation inside each zone. The flows involved in this model are coupled with the thermal system, which enables simultaneously taking into account all the different thermal transfers.

The humidity model leads to a system of equations similar to the thermal model and does not take into account the humidity transfers in the partitions and the furniture.

*3.2 Definition of a typical day*

Reunion is situated at a latitude of 21° South and a longitude of 55° East. The climate is humid tropical.

There is a dry season (May to October) predominated by the trade winds, mainly cold and dry and a wet season (November to April), hotter and more humid with light winds from differing directions. The island is also influenced by the passage of cyclones.

The relief splits the island in half :

- The windward zone, exposed to the trade winds, characterised by heavy rainfall and an average temperature of 23°C.
- The leeward zone, generally sunny and dry. This region has much less rainfall and the average temperature is 2 to 3 degrees higher than the first zone.

The study of the effects on a building from outside conditions, necessitates the availability of certain meteorological informations, representative of the studied site. In order to reduce the calculation time, our simulations were carried out over one day only. This day had to be that which was most representative of the conditions of the wet season. It is in that period that the most unfavourable combination of parameters are found.

The day which most closely represented the average temperature and solar radiation conditions observed on site was selected. Also, in order to optimise the radiative gain of direct solar radiations, conditions with low cloud covering were chosen. (see Fig. 4).

The choice of site for the selected meteorological sequence was the weather station at Gillot, situated in the North East of the island, not far from St Denis. It is a highly populated urban zone. The site is representative of a humid coastline, at least during the wet season, when the trade winds are not blowing.

Table 1 gives the values found for each criteria for the given site :

*Table 1. Meteorological criteria used for the typical day*

Fig. 4 : *Typical day used in the simulations*

*Wind :* Concerning the wind speed and wind direction, our aim was to simulate the thermal behaviour of the dwelling during the warm humid season, which is a period where the trade winds do not occur.The Reunion Island is influenced by the phenomena of land-sea heating and cooling effects, creating thermal winds (onshore during the day, offshore during the night). Following the consultation of the meteorological figures in our posession, it was found that a night breeze which speed attains 1m/s, represents 20% of the distribution in frequency and that its direction is perpendicular to the coast line at night.

We also created an artificial wind file, consisting of seven consecutive days with the same external temperature, relative humidity and solar radiation characteristics (the same as the typical day above) but with different wind speeds and wind directions. We considered the case of days with a light wind ($v = 1$ m.s$^{-1}$), and days with a moderate wind ($v=5$ m.s$^{-1}$) from variable directions. This way it was possible to assess the airflow performances of the dwellings in different wind situations.

Day 1 : no wind

Day 2 : light wind (1 m.s$^{-1}$), South East

Day 3 : moderate wind (5 m.s$^{-1}$), South East

Day 4 : light wind, East

Day 5 : moderate wind, East

Day 6 : light wind, North

Day 7 : moderate wind, North.

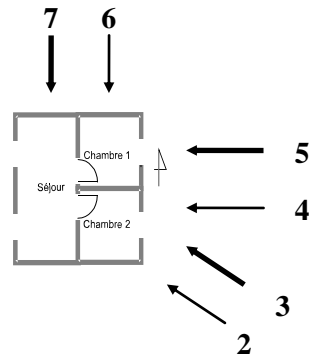

*3.3 Description of a typical dwelling :*

The typical dwelling chosen is that which is most reprsentative of the type of accomodation built in the Reunion Island, in terms of architecture and building materials. We have selected two typical dwellings which conform to the plans figure 6 but differ on a level of thermal inertia (one is a light structure, the other a heavy structure ) and a dwelling in block of flats. These dwellings will constitute the references for our simulations. Table 2 summarizes all the components of each dwelling.

- *The typical individual dwelling*

*Fig. 6 : Vues of the typical individual dwelling*

This dwelling is composed of three bedrooms, a living room, a bathroom and a kitchen.

We considered only the living zones of the house for our simulations, bedrooms 1 and 2 and the living room, for multiple reasons:

1) The standard concentrates exclusively on the improvment of the living areas

2) Taking into account the number of aspects covered we have tried to optimise the number of zones, that is why a three roomed building (Fig. 7) seemed the most obvious choice to highlight the different thermal, air flow, influence of orientations, etc ...phenomena.

The glazed surfaces in the rooms represent 11% of the total surface, and in the living room 22%. The bedroom surfaces are 11 m2 with a volume of 30 m3, the living room surface is 22 m2 with a volume of 60 m3.

- *Flats :*

After consulting the local low cost housing institutions and the new housing statistics for Reunion Island, we found the most common dwelling found is the T3/V, consisting of two bedrooms, a living room and a veranda (see Fig.8). The simulations were carried out on three types of dwelling (one beneath the roof, one between two stairs, and the last one on the side of the building), which are part of the building elements classification of De Waal [7] (see Fig. 9). For our simulations, we assumed the typical flat to be a two zone building (Living room zone and bedroom zone).

*Table 2 : Description of typical dwellings*

*Fig. 8 : Typical flat*

*Fig. 9: Case studies*

*3.4 Comparison criteria:*

To compare the different technical solutions, the thermal comfort aspect as well as the energetic aspect were taken into account.

*Comfort criteria:*

The comfort parameter we chose is the resultant temperature. This variable enables us to take into account the non-comfort arising from the long wave radiative effects.

$$Tres = 0.55 \cdot Ta + 0.45 \cdot Tr \qquad (2)$$

where   Ta : indoor air temperature (°C)

Tr : mean radiant temperature :

$$Tr = \frac{\sum_{i=1}^{n} S_i \cdot Tsi_i}{\sum_{i=1}^{n} S_i} \qquad (3)$$

where $Ts_i$ is the indoor surface température for zone i, and Si the i component surface.

We followed the evolution of this variable along our typical day but we used average day and night temperatures, characterizing the day and night uses of the dwelling.

- day resultant temperature : average resultant temperature from 7h00 to 19h00
- night resultant temperature : average resultant temperature from 20h00 to 06h00
- maximum resultant temperature

*Energetic criteria :*

Another important criteria is the energetic criteria. We have used this criteria to measure power and energy improvements brought about by the technical solutions. The criteria used in our simulations are:

- maximum power (W)
- maximum power per square metre (W/m$^2$)
- total daily thermal energy(thermal kWh)
- overall thermal energy during the wet season (thermal kWh)

The power observations were carried out on the base of an infinate power air conditionning system keeping each room temperature at 25°C. This method thus enabled us to obtain results in energy consumption and in thermal power reduction.

*3.5 The simulation strategy:*

*The components :*

As far as outer protecting structures are concerned (ie. roof, walls, windows), we have carried out a temperature and power study for each dwelling and each component part thereof. During our simulations, the building is closed and without air renewal. Only the thermal performance aspect is taking into consideration. This initial phase enables us to obtain the best technical solutions. For example, when considering an opaque separation, do we need a solar protection of d/h = 0.25, 0.50, 0.75 or 1 (d : dimension of the over-hang and h : height of the wall), or is insulation better ? We have therefore tested different roof colours with differing thicknesses of insulation varying from 0 cm to 10 cm, different opaque separation colours with a solar protection in the form of an over-hang varying from d/h = 0 to d/h = 1, or by thermal insulation to which the thickness varies from 0 cm to 4cm. As regards the solar protection for the glazing, we compared different dimensions of horizontal external sunshades and venitian blinds.

Natural ventilation is a case appart, as we may only judge the improvments by the temperature reached. The problem posed was to measure the impact of external and internal permeability on the internal resulting temperature. In brief, what percentage of facade openings should one have to improve thermal confort, and does interior lay out have any influence on this ? The bioclimatic perfectionists site percentages of 40%, which are to high to be economically possible [17]. We have varied the interior and exterior permeability rates of typical dwellings from 15% to 40% and have simulated all the possible combinations.

*Air conditioned room option*

We suppose that one of the rooms is air conditioned, with a cold production period from 20h00 to 6h00. The internal gains are constituted by four people (2 adults et 2 children), and by lighting in each room. The following table summarizes the internal charges .

*Table 3 : Internal charges summary table - individual dwelling*

The simulations were carried out on the base of a bad thermal structure building, an important air renewal rate (5 vol/h) and a set temperature of 26°C and 22°C and a dwelling with good solar protection, a controled air renewal rate of 1 vol/h and a set temperature of 26°C.

*Real case:*

With all the technical solutions of each component of the outer structure conformed to, we compared, in a second phase, the existing dwelling which is in overall badly designed (bad solar protection, insufficient

ventilation,etc...), to a well designed dwelling adhering to the technical solutions that we had found during the first phase.

## 4. RESULTS

*4.1 Location on site:*

Performant thermal and energetic housing conception starts immediately from their location on the building site. The immediate surroundings of a building have a significant influence on the conditions of thermal comfort inside. This is particularly the case of the surrounding surface of the building, which should neither reflect the solar radiations towards the house nor increase the ambient air temperature.

The results concerning the surroundings are :

The finished surface around the building should be protected from direct sunlight on more than three quarters of its perimeter, with a width of at least 3 metres. This can be satisfied by either vegetation (lawn, bushes, flowers) around the building, or by all vegetation sun-blocks. These prescriptions are similar to the recommendations of De Wall [7] concerning urban planning for warm humid climates.

*4.2 Solar protection:*

In a humid tropical climate, the sources of uncomfort arise from a temperature increase due to bad architectural conception, concerning insulation. 80% of this is due to solar radiation, the rest, to conduction exchanges. The setting up of an efficient solar protection constitutes the second fondamental phase in the building thermal conception. This protection concerns all the exterior separations of the dwelling : roof, walls and windows.

*Solar protection of the roof :*

Thermal inflows represents up to 60% of the overall inflows from the separations in the dwellings. An efficient solar protection for the roof is therefore of prime urgence for a good thermal conception.

The following table is valid for terrasse rooves, inclined rooves without lofts, rooves with closed or barely ventilated lofts.

When concerning well ventilated lofts, these should have ventilation ducts spread out uniformly through all its perimeter and which surface conforms to the following inequation :

$$\frac{S_0}{S_t} = \frac{Total\ area\ of\ openings}{Roof\ area} \geq 0.15 \qquad (4)$$

In this case, the ceiling under the loft should satisfy certain prescriptions (see table 4).

*Table 4 : Roof solar protection*

In general using a light colour is the only way in which we are able to lower the inside temperature of the dwelling by 3°C, equally, insulation adds a suplimentary decrease of 3°C. The avoided thermal power is 150W per m² of protected area, and the avoided thermal energy is equal to 250kWh/m² for the whole of the wet season.

Equally it is possible to use thermal insulation barriers such as aluminium foil-coated products [18]. They must be installed in attics, with an adjacent airgap. The aluminium foil reflects radiant heat like a mirror, whilst the polished aluminium emits little of the radiant heat that falls on it. This type of protection is very efficient in climates with high levels of sunshine.

*Solar protection of walls :*

The thermal gains from the walls represents 20 to 30% (40 à 65 % for the dwellings which are not under the roof) of the thermal gains from the seperations. Various solutions enable a protection of the walls from the sunlight : horizontal or vertical canopy or overhang, thermal insulation of the walls.The results obtained from the simulations constitute the following table, which give the optimum dimensions of the canopy in relation to the orientation of the walls and the walls inertia.

*Table5 : over hang of canopy P- minimum d/h ratio values to be respected.*

When the walls do not have a canopy, the minimum thicknesses of insulation (in cm) needed for the different types of walls and in different orientations are shown in the following table :

*Table 6 : Insulation of walls (in cm) for different orientations and external colours (for a conductivity of 0.041 W/m.K)*

These solutions lead to a reduction in the resulting interior temperature of 0.5°C (heavy structure) to 1°C (light structure). They enable a reduction of the entering thermal flows of 40W per m² of protected area, corresponding to an avoided thermal energy of 65kWh/m² for a light structure. For a heavy structure, the reduction of entering flow is 15 W/m² and the avoided thermal energy is 25 kWh/m².

*Solar protection of windows :*

The protection of the windows is fundamental, not only because they represent 15 to 30% of the thermal gains but also because they contribute to the increase in the uncomfort experienced by the occupant, due to the instant heating of the ambiant air temperature and an exposion to direct or reflected sunlight. All the windows must therefore be protected by some sort of window shading, such as horizontal canopies and other shading devices such as venitian blinds or opaque, mobile strips.

The simulations enabled us to optimize the geometric characteristics of the horizontal canopies in relation to the orientation of the glazing. (see table 7).

*Table 7 : Values of d/(2a+h) (case 1), or d/h (case 2)*

These solutions lead to a reduction in the interior temperature of more than 4°C for a light structure and 2°C for a heavy struture.They enable the reduction of the thermal flows of 120W per m² of protected window for a light struture, corresponding to an avoided thermal energy of 130 kWh/m². The reduction of the thermal flow for a heavy structure is of 100 W/m² and the avoided thermal energy is 100 kWh/m².

*4.3 Natural ventilation :*

In warm climates, natural ventilation is the most usual means of heat transfer from both occupants and buildings [18].

The natural ventilation ,depending on its importance, ensures three functions [7], [17].

- *Weak flow (1 to 2 vol/h)* for the preservation of hygiene conditions by air renewal.
- *Moderate flow (40 vol/h),* for the evacuation of internal gains and the cooling of the outside structure.
- *High flow (more than 100 vol/h)* to assure the comfort by sudation. Thus the high air speed and its good layout betters the sudation process. This is the only means which enables the compensation of the high temperatures, coupled to a high rate of hygrometry.

Our aim is therefore to find the exterior/interior permeability coupling which enables us to obtain the rate of air renewal of 40 vol/h. On the one hand the structure of the dwelling will be sufficiently cooled and on the other, such an air renewal rate allows us to hope for wind speeds of 0.2 à 0.5 $m.s^{-1}$ , which is largely sufficient, when taking into account the climatic parameters (outside temperature rarely greater than 32°C), to assure a good level of comfort.

We found from the simulations that the critical air renewal rate of 40 vol/h is obtained for a configuration of exterior permeability equal to 25% and interior permeability of 25%, equally for a light structure, as for a heavy structure. The natural ventilation is simply more effective during the night for the heavy structure, whereas in the light structure it serves mainly to evacuate the overheating from during the day (see Fig. 10).

Thus the dwelling should have complete cross ventilation (see Fig. 11). At each level or floor, there should exist openings in the principal rooms, on at least two opposing facades (the principal rooms being the bedroom and the living room).Also the interior lay-out should be designed in a way that the outside air, flows through the principal rooms and the corridoors, from one facade to the other, by the doors and the other openings in the partitions.

*Fig 11 : Cross ventilated dwelling*

$$P1 = \frac{So1}{Sp} \geq 0.25$$

$$P2 = \frac{So}{Sp} \geq 0.25 \qquad \text{(5a, b, c, d ,e)}$$

$$Sp = \frac{Sp1 + Sp2}{2}$$

$$Si1 \geq So1 \text{ or } So2$$

$$Si2 \geq So1 \text{ or } So2$$

So1 : Net surface of openings, principal rooms (façade 1).

So2 : Net surface of openings, principal rooms (façade 2)

Sp1, Sp2 : Total surface of principal rooms of façades 1 and 2.

Fig.11 gives the details of the calculations needed to determine the exterior and interior permeabilities of 25%.

*Air fans :*

When natural ventilation air speed is unsufficient, air fans can additionally be used. This enables an increase in the comfort range of more than 2°C [19].
Each room in the dwelling should be equipped with electric wiring in the ceiling, wired to a wall switch, destined exclusively for the installation of air fans.

*4.4  Air conditioned bedroom option:*

In certain dwellings, and at certain times of the year, natural ventilation, even with the existance of air fans, is not adequate for an acceptable level of comfort. In this case we can choose to air condition the bedrooms using efficient appliances. The simulations which we carried out show that the air conditioning charges can be reduced through good structure conception and control of the air renewal rate. For a light structure these savings reach 3.4 cooling kWh per day, and 11 kWh for a heavy structure, where the inertia plays a dominating role in the air conditioners consumption. Throughout the whole of the wet season, the consumption was diminished by half with a good structure conception (1000 cooling kWh). The maximum cooling power is therefore 80W/m².

Practically, the air conditioners should meet certain standards of efficiency (cooling efficiency of 2.5 for the window units and 3 for split-systems), of permeability (each room should be equipped with a mechanically controlled air renewal of 25m$^3$/h) and a maintenance contract.

*4.5 Domestic hot water*

We have not carried out simulations in this domain, however a certain number of prescriptions need to be verified, and we feel that they need to be specified as water heating consumes greatly and constitutes a real energy problem (see part one).It is important that the dwellings are equipped with efficient long-lasting and economic, domestic hot water heating systems.The water heater can be solar, electric or gas.

In the case of solar water heaters, the apparatus must conform to the technical control CSTB.The total minimum surface of the solar captors, should be installed in relation to the size of the dwelling.(see table 8). The capacity of the water storage should be 60 and 120 litres per square metre per net square metre of captor.The conventional minimum annual production should be 700kWh per net square metre of the capting surface.

*Table 8 : Technical characteristics - solar water heaters*

When considering electrical heaters the appliance must have the mark of the approved French standard of manufacture NF (Norme Française). The instant hot water heaters are high energy consuming, so therefore are excluded.The capacity of the water heater and its.cooling constant are calculated in function to the number of principal rooms within the dwelling (see Table 9).. The power supply should be equipped with a three position commutation swich: servo-controlled to the off peak hours, over-ride, off.

*Table 9 : Technical characteristics of electric water heaters*

The gas water heaters must have the French Standard Mark and also provision must be made for a burnt gas outlet.

*4.6 Case study*

In this part of the paper we have compared the basic, badly designed dwellings with the dwellings which respect the prescriptions of the Standard in terms of solar protection and natural ventilation. Fig. 12 to 15 show the comfort gains in the living room area for an individual light structure and a heavy structure. With good thermal conception, the resulting inside temperature of the dwelling remains inferior to the outside temperature.

*Fig. 12 :Case study - light structure*

*Fig. 13 :Case study - heavy structure*

*Fig. 14 :Comfort gains - light structure*

*Fig. 15 :Comfort gains - heavy structure*

We can also show evidence that each technical solution improves the interior comfort. (see Table10). We found that more than 70% of the comfort improvment comes directly from the solar protection of the roof and the natural ventilation, therefore these are the essential components of comfort improvment.

*Table 10 : Percentage change in comfort due to each passive cooling strategy*

## 5. CONCLUSION

The simulations carried out have led to the definition of performant passive technical solutions for each comprising part of the structure and likewise a minimum ratio to optimise the natural ventilation. The dwellings to be constructed following ECODOM specifications should satisfy the criteria of these technical solutions.

A technical-economical study was launched to evaluate the excess cost of materials. In an overall operation, the excess cost does not exceed 10 000 French Francs (1500 EURO). The aim is therefore to provide the owner with an incentive by the means of a subsidy of approximately 6000 FF (1000 EURO) per standardized dwelling.

The objective, in terms of number of dwellings, is in the order of 6000 dwellings within three years. The total energy gain evaluated is 6 MW and 23 Gwh electric (5100 TEP). There will be important repercussions on the environment as 23 Gwh of electricity economized represents at first approach, a reduction in emissions of 5000 T of $CO_2$.

An experimental follow up of the dwellings will be organised for the first ECODOM operations, in order to validate experimentally the impact of the passive cooling solutions on the comfort of the occupants. This follow up is important, as the setting up of The ECODOM Standard will be the first step towards the setting up of thermal regulations in the French overseas departments, by the year 2000.

**Acknowledgement :** The contribution of Electricité de France Ile de la Réunion to this study is gratefully acknowledged.